\documentclass[aps,prl,twocolumn,showpacs,floatfix,superscriptaddress]{revtex4}
\usepackage{graphicx}
\usepackage{amsmath}
\usepackage{amssymb}
\bibstyle{apsrev.bib}

\newcommand{\beq}{\begin{equation}}
\newcommand{\eeq}{\end{equation}}

\begin{document}

\title{Intensity Thresholds and the Statistics of the Temporal
  Occurrence of Solar Flares}
 
\author{Marco Baiesi}
\affiliation{INFM-Dipartimento di Fisica, Universit\`a di Padova,
  I-35131 Padova, Italy.} \affiliation{Sezione INFN, Universit\`a di
  Padova, I-35131 Padova, Italy.}
\affiliation{Instituut voor Theoretische Fysica, K.U.Leuven, Belgium.}

 \author{Maya Paczuski} 
\affiliation{John-von-Neumann Institute for Computing, 
Forschungszentrum J\"{u}lich, D-52425 J\"{u}lich, Germany.}
\affiliation{Perimeter Institute for Theoretical Physics, 31 Caroline St. North, Waterloo, Ontario 
N2L 2Y5 Canada}

\author{Attilio L. Stella}
\affiliation{INFM-Dipartimento di Fisica,
Universit\`a di Padova,  I-35131 Padova, Italy.}
\affiliation{Sezione INFN, Universit\`a di
Padova, I-35131 Padova, Italy.}  
\date{\today}

\begin{abstract}
  Introducing thresholds to analyze time series of emission from the
  Sun enables a new and simple definition of solar flare events, and
  their interoccurrence times.  Rescaling time by the rate of events,
  the waiting and quiet time distributions both conform to scaling
  functions that are independent of the intensity threshold over a
  wide range. The scaling functions are well described by a two
  parameter function, with parameters that depend on the phase of the
  solar cycle.  For flares identified according to the current,
  standard definition, similar behavior is found.
\end{abstract}


\maketitle

The solar corona is a very high Reynolds number turbulent plasma
producing intermittent bursts of radiation.  Plasma forces twist the
coronal magnetic fields until stresses are suddenly released, an
avalanching process governed by magnetic reconnection~\cite{lh}.  The
released magnetic energy induces radiative emission that can be
detected as a flare. Flares exhibit scale invariant statistics.  
For instance, the probability distribution of flare energies is a power
law spanning more than eight orders of
magnitude~\cite{aschwanden:TRACE,dennis}, similar to the
Gutenberg-Richter law for earthquakes.  The
distribution of magnetic concentration sizes on the photosphere is
also scale invariant, and the coronal magnetic network embodies a
scale-free network~\cite{hughes_pac1,pac_hughes1}.  In fact, a model of
self-organized criticality (SOC) with   avalanches of reconnecting flux tubes reproduces the
observed scale-free network structure~\cite{hughes_pac1,hughes_pac2}.

As part of the debate on the characterization of magnetohydrodynamic
turbulence in this
regime~\cite{hughes_pac1,pac_hughes1,hughes_pac2,lh,Wheatland:Waiting,
Boffetta:Waiting,GROUP1}, interest has focused
on comparing interoccurrence times between flares with those in models
of SOC.  Analyses of flare catalogs have
indicated scale invariance of the waiting times, but the behavior was
found to vary with the phase of the solar
cycle~\cite{wheatland02:_minmax} and with the methods used to analyze
the catalogs.  (See e.g.~Ref.~\cite{wheatland02:_minmax,gls}.)  The prior
belief that avalanches occur with Poissonian waiting times in
the well-known BTW sandpile model~\cite{btw} (giving an exponential distribution of
waiting times) argued against the SOC
hypothesis~\cite{Boffetta:Waiting}.  However, 
including a finite detection threshold leads to a power law distribution 
of quiet times even for the BTW model~\cite{pbb}, when
durations and quiet times are measured with the same clock.  Since the
turnover time scale for flux to be regenerated in the corona is of the
order of ten hours~\cite{hagenaar2003}, while the correlated waiting
time intervals between flares can extend up to years, the physical
mechanism(s) responsible for these correlations resides in
the turbulent convective region beneath the photosphere that generates
magnetic flux and drives it into the corona.  Systematic studies of
the temporal pattern of flares can give insight into the dynamics of
magnetic flux in the convective region, or the solar dynamo,
which is difficult to observe directly.

Here we show that the interoccurrence times between flares has a
hierarchical scaling structure when flares are defined as intervals
during which the emission exceeds a given threshold. Rescaling time by
the rate of these events, we find universal behavior for the
interoccurrence times, which is independent of threshold. 
Both at solar maximum and at solar minimum the scaling function
can be fitted  by a simple two-parameter
function. This generalized Lorentzian arises naturally within a simplified
model based on the time-dependent Poisson process. From this model
one can naturally infer an exponential distribution of flaring rates
at solar maximum. At solar minimum the distribution of quiet- or laminar -
times is accurately described by on-off intermittency~\cite{on-off}, 
a mechanism already proposed to describe the solar cycle~\cite{dynamo}.

\begin{figure}[!tb]
\includegraphics[angle=0,width=8.0cm]{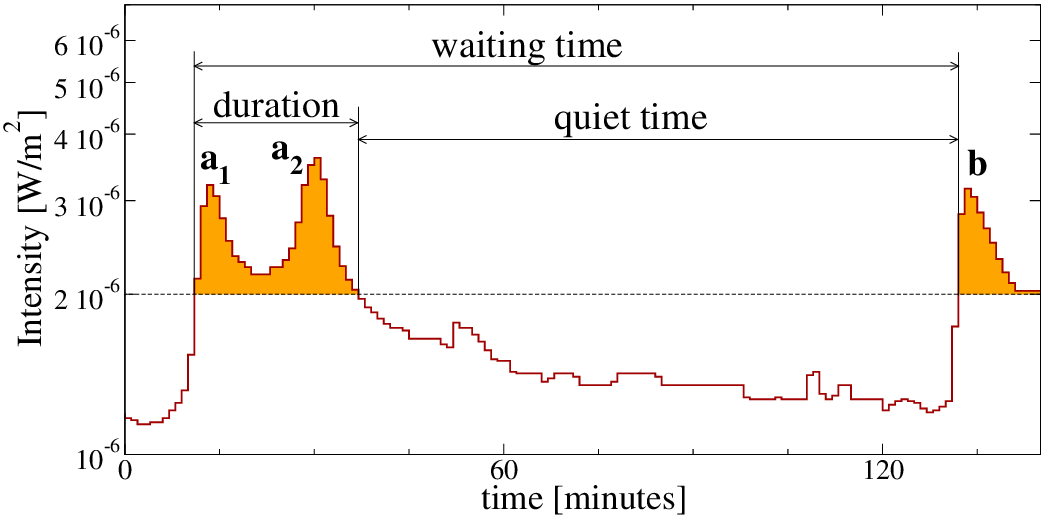}
\caption{(Color online) 
Explanation of various times used in this work.
Here the threshold intensity is $I=2\times 10^{-6}$  W/m$^2$.
According to the standard definition of flares, ${\bf a_1}$ and ${\bf a_2}$ 
would be two separate events.
In our case they are separated if, for instance, 
$I=3\times 10^{-6}$ W/m$^2$.
This shows that the set of events defined by different thresholds are not
trivially related to each other or to the
flares listed in the standard catalogs.}
\label{fig:ex}
\end{figure}

In extremely intermittent time series, like e.g. earthquakes, events
are spikes separated by a smooth background, and are easily and
uniquely defined.  This is not the case for the solar data analyzed
here, where the intensity decays slowly after a local peak, allowing
overlaps with subsequent peaks.  In this case,
the introduction of a threshold is deeply connected to the definition
of events, as indicated in Fig.~\ref{fig:ex}.

%%%%%%%% FIG  %%%%%%%%%%%%%%%%%%%%%%%%%%%%%%%%%%%%%%%%%%%%%%%%%%%%%%%%%
%%%%%%%% FIG  %%%%%%%%%%%%%%%%%%%%%%%%%%%%%%%%%%%%%%%%%%%%%%%%%%%%%%%%%

The time series have been downloaded from the ``Space Physics
Interactive Data Resource'' (SPIDR) web site~\cite{NOAAdata}, where
each bin represents X-ray flux averaged over a given time unit.
Among the available signals from various GOES satellites~\cite{NOAAdata}, 
we consider the
time series of the average soft X-ray flux  measured in
W/m$^2$ with photons in the
range from 1 to 8 \AA . See Table~\ref{tab:1} for details. 
We have also isolated two periods, roughly corresponding to the 
most recent minimum (``min'') and maximum (``max'') of the solar cycle.
For comparison, we also analyze data in publicly available
flare catalogs~\cite{NOAAdata_Hughes}, which identify events with time
intervals when the intensity is higher than a local average of the
signal. 

In contrast, we define an event, or flare, to be simply the interval
during which the intensity exceeds a certain fixed threshold.  To
compensate for fixing the threshold we study, in detail, the
dependence on the threshold value and obtain results that are
independent of the threshold over a wide range.  Various times related
to our definition of events (durations, waiting and quiet times) are
explained in Fig.~\ref{fig:ex}.  One could expect differences when events defined
by our simple procedure or by the standard flare catalogs are
analyzed.  However, we find that the statistical distributions are
mostly similar.
Hence, we expect similar results if other criteria are used to define 
interoccurence times, such as the time
difference between subsequent maxima in the signal, referred to as the 
"laminar times" in Ref.~\cite{Boffetta:Waiting}. Indeed, that definition
also allows a systematic variation of intensity threshold used to select 
maxima that could be compared with the results shown here.

\begin{figure}[!tb]
\includegraphics[angle=0,width=8.0cm]{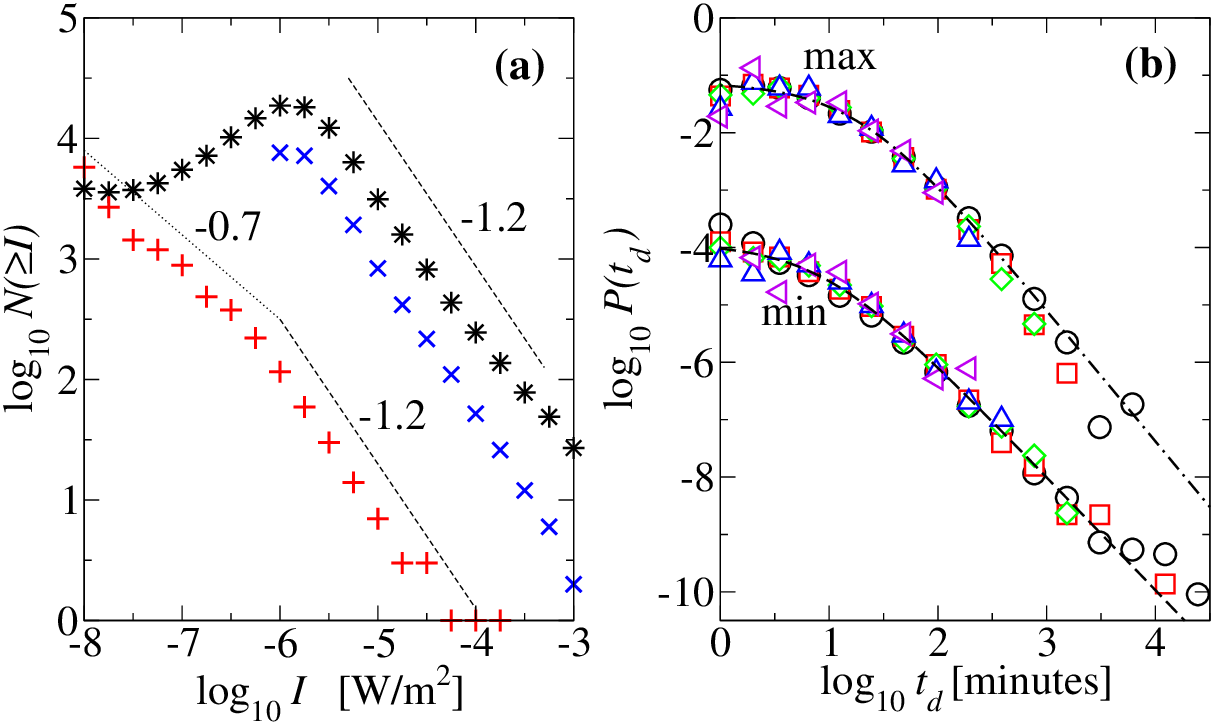}
\caption{(Color online) (a) Number of events
with intensity greater than a given threshold, for the entire record
$(\star)$, at the minimum $(+)$ and maximum $(\times)$ of the cycle.
The straight lines represent power laws with the quoted exponent.
(b) The distribution of flare durations, for different thresholds,
at ``min'' and ``max'' of the solar cycle 
(the former shifted down by three units on the log-scale). 
The  symbols are explained in Table~\ref{tab:1}, while the curves are
fits using Eq.~(\ref{eq:q-exp}). 
\label{fig:PId}}
\end{figure}

The number of events with intensity greater than or equal to a given
threshold $I$, $N( {\ge} I)$, is shown in Fig.~\ref{fig:PId}(a) for the entire
data set,  at ``min'' and at ``max''.
For the whole catalog the number of flares $N({\ge} I)$
behaves approximately as $I^{-\beta}$, with $\beta= 1.2(1)$ for
intensities greater than $\approx 10^{-6}$~W/m$^2$.  Scaling
breaks down below $I\approx 10^{-6}$, where $N({\ge} I)$ increases
with $I$.  This clearly shows that $N( {\ge} I)$ is not the cumulative
version of any probability distribution, because flares at different
thresholds are different objects.  In the two
sub-regimes ``min'' and ``max'', we find power law behavior $N( {\ge}
I) \sim I^{-{\beta_{\max}}}$, with ${\beta_{\max}}= 1.2(1)$, for
sufficiently large $I$.  During the minimum of the cycle, however,
another scaling regime appears.  Indeed, $N_{\min}( {\ge} I) \sim
I^{-{\beta_{\min}}}$, with ${\beta_{\min}}= 0.7(1)$, for $I\lesssim
10^{-6}$~W/m$^2$.  Within statistical error, the exponent
$\beta_{\max}$ agrees with the (cumulative) distribution of peak fluxes
measured by Aschwanden {\it et al}~\cite{aschwanden:TRACE},  who
obtained $\beta_{\max} +1=2.08 \pm 0.03$.

%%%%%%%% FIG  %%%%%%%%%%%%%%%%%%%%%%%%%%%%%%%%%%%%%%%%%%%%%%%%%%%%%%%%%
%%%%%%%% FIG  %%%%%%%%%%%%%%%%%%%%%%%%%%%%%%%%%%%%%%%%%%%%%%%%%%%%%%%%%

%%%%%%%%%%%%%%%%%%%%%%%%%%%%%%%%%%% _________ TABLE __________
\begin{table}[!b]
\caption{\label{tab:1} 
Time series and  selected intensity thresholds.}
\begin{ruledtabular}
\begin{tabular}{l|lll|c}
\ &
All \footnotemark[1]&
min.\footnotemark[2]& 
max.\footnotemark[3]& 
sym.\footnotemark[4]\\
\hline
start& 1/1/'86&1/9/'95&1/1/'00& \\
end& 31/3/'04&31/12/'96&31/12/'03& \\
bin width (minutes)& 5  & 1  & 1  & \\
\hline
thresholds (W/m$^2$)& $2{\times} 10^{-6}$ &  $3{\times}10^{-8}$ & $2{\times} 10^{-6}$&{\Large{$\circ$}}\\
& $4{\times}10^{-6}$& $10^{-7}$& $4{\times}10^{-6}$&$\square$\\
& $10^{-5}$&$3{\times}10^{-7}$&$10^{-5}$&{\Large{$\diamond$}}\\
& $3{\times}10^{-5}$&  $10^{-6}$& $3{\times}10^{-5}$&$\triangle$ \\
& $10^{-4}$&$3{\times}10^{-6}$&$10^{-4}$&$\vartriangleleft$
\end{tabular}
\end{ruledtabular}
\footnotetext[1]{By concatenating signals of satellites
GOES 6, 7, 8, and 10~\protect\cite{NOAAdata}, 
we reconstruct the time series representing almost two solar cycles.
Since some data are missing, values in the empty bins are set by
the last recorded value before each of them. In this way, flat
plateaus of  intensity are created, 
introducing a minimal bias into the data.}
\footnotetext[2]{Data from GOES-8, minimum of the solar cycle.}
\footnotetext[3]{Data from GOES-10, maximum of the solar cycle.}
\footnotetext[4]{Symbol used in the figures to denote the corresponding data.}
\end{table}
%%%%%%%%%%%%%%%%%%%%%%%%%%%%%%%%%%% _________ TABLE __(end)___

The thresholds and associated symbols used by us to define the events
are given in Table~\ref{tab:1}.  For the entire data record, we choose
five thresholds with $I>10^{-6}$, where $N({\ge} I)$ is a decreasing
function of $I$. The same thresholds are also used in the ``max''
regime.  Since the flux at the maximum of solar activity is typically two
orders of magnitude greater than at the minimum,  a definition of
flares by means of the same set of thresholds could be
unfeasible. Five different thresholds are used at solar
minimum to obtain reasonably good statistics.

We first discuss the distribution of duration times, $P(t_d)$. It has
a power law tail, with critical exponent $\gamma_{\rm dur}$, which extends to
longer durations on lowering the threshold, as shown in
Fig.~\ref{fig:PId}(b).  Previous measurements using a different definition of
flares than that put forward here found 
$\gamma_{\rm dur} = 2.17$ to $\gamma_{\rm dur}=2.54$,
depending on the range of times fitted \cite{Crosby:FlareObsXB}, while
Litvinenko obtained  $\gamma_{\rm dur}=2$ using dimensional analysis~\cite{litvinenko}.  

However, $P(t_d)$ crosses over from a power law at large times to a
constant regime at short times.  The entire distribution for all thresholds and all time
periods
is consistent with the function
\beq P(t_d) \sim (1 + t_d / t_d^*) ^ {-\gamma_{\rm dur}}
\label{eq:q-exp}
\eeq 
with $\gamma_{\rm dur}=2.0(1)$ \cite{note1}, and $t_d^* \approx 10$ min at solar
minimum while $\gamma_{\rm dur}=2.3(4)$ and $t_d^* \approx 20$ min at solar maximum~\cite{note_fit}.

%%%%%%%% FIG  %%%%%%%%%%%%%%%%%%%%%%%%%%%%%%%%%%%%%%%%%%%%%%%%%%%%%%%%%
\begin{figure}[!tb]
\includegraphics[angle=0,width=8cm]{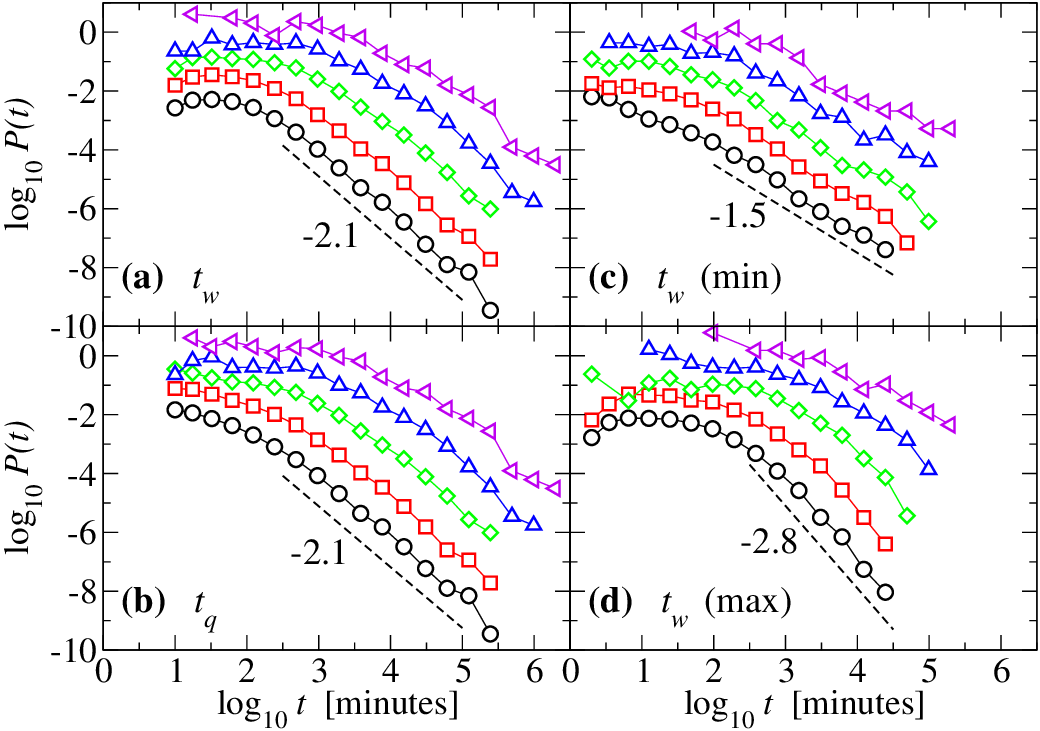}
\caption{(Color online)
The distributions of recurrence times, progressively shifted along the $y$-axes
with the threshold, for clarity. 
(a), (c) and (d): Distributions of waiting times, for the entire
record and at the minimum and maximum of the solar cycle, respectively.
(b): Distribution of quiet times for the entire record. The 
power law tails are indicated with straight lines whose slope is $\gamma$.
The symbols are explained in Table~\ref{tab:1}.
\label{fig:PwPq}}
\end{figure}
%%%%%%%% FIG  %%%%%%%%%%%%%%%%%%%%%%%%%%%%%%%%%%%%%%%%%%%%%%%%%%%%%%%%%

The waiting and quiet time distributions are shown in
Fig.~\ref{fig:PwPq} for different thresholds and regimes.  Each
$P(t_w)$ is similar to its respective $P(t_q)$, especially at large
times, where both decay as power laws $\sim t^{-\gamma}$. 
Hence, the scale-free duration of flares is not giving
peculiarities in the passage from $t_q$ statistics to $t_w$ ones.  The
waiting/quiet time exponents $\gamma$ at lower thresholds have been
evaluated in the three regimes. Within statistical error these values,
indicated in Fig.~\ref{fig:PwPq}, agree with the
ones determined by Wheatland and
Litvinenko~\cite{wheatland02:_minmax}, who analyzed flare catalogs.
  However, one can observe
that by increasing the threshold, both $P(t_w)$ and $P(t_q)$ evolve
continuously, becoming flatter up to longer times for higher
thresholds.  This aspect was not caught by any previous 
studies~\cite{wheatland02:_minmax}, 
whose results were obtained without systematically varying any threshold.

\begin{figure}[!tb]
\includegraphics[angle=0,width=7.1cm]{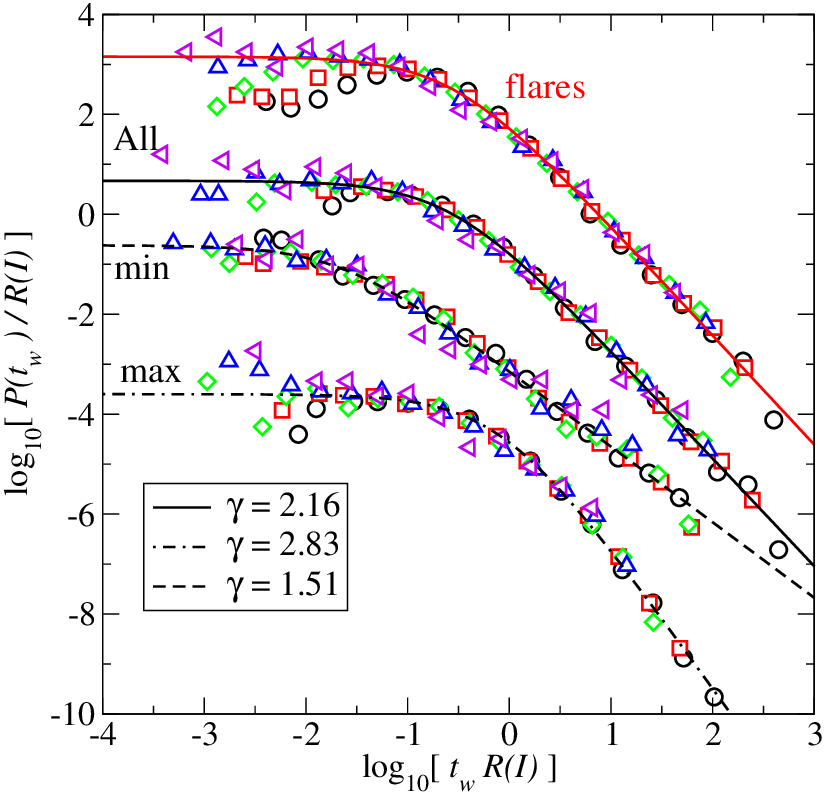}
\caption{(Color online) 
  The rescaled distributions of waiting times,
  arbitrarily shifted vertically to be distinguished. From below,
  for the maximum of the solar cycle, for the  minimum, 
  for the entire catalog, and
  for the GOES flare catalog.  
  Fits as described in the text 
  are shown for the different phases.
\label{fig:Pw_resc}}
\end{figure}

A scaling argument similar to one recently put forward by Bak {\it et al}
for waiting time statistics of earthquakes~\cite{bak02:_unified}
can unify in a single scaling function the waiting/quiet time statistics.
We argue that $N( {\ge} I)$ provides the right rescaling factor for
the recurrence times, namely, the one that gives a collapse onto a single
curve of all the
distributions measured with different intensity thresholds, $I$.
In particular, we rescale the interoccurrence
times by their average, which is inversely proportional to the average 
rate of events, $R(I) = N( {\ge} I) / \Delta T$~\cite{corral03:_unified}, 
where $\Delta T$ is the time span of the record. 
Thus, the distribution of waiting times for a given threshold is given by 
\beq 
P(t_w,I) \sim R(I)g(t_w R(I)) \quad . \label{eq:2}
\eeq   
Unlike the universal waiting time distributions
for earthquakes~\cite{bak02:_unified,corral03:_unified}, but similar
to $P(t_d)$, the scaling function $g$ for the flare waiting and quiet
times is also well-described by the function, $g(x)\sim (1 + x/x^*)^{-\gamma}$, 
as shown in Fig.~\ref{fig:Pw_resc}.
Furthermore, the rescaled distributions have parameters 
that depend on the phase of the solar cycle: for the whole catalog,
$\gamma\simeq 2.16(5)$ and $x^* \approx 0.26$.  At the
minimum of the cycle $\gamma=1.51(5)$ ($x^* \approx 0.02$), while at
the maximum $\gamma=2.83(10)$ ($x^* \approx 0.85$).

%%%%%%%% FIG  %%%%%%%%%%%%%%%%%%%%%%%%%%%%%%%%%%%%%%%%%%%%%%%%%%%%%%%%%
%%%%%%%% FIG  %%%%%%%%%%%%%%%%%%%%%%%%%%%%%%%%%%%%%%%%%%%%%%%%%%%%%%%%%

To compare with the standard definition of flares, we repeat the above
analysis using the GOES flare catalog from
1975-2003~\cite{NOAAdata_Hughes}.  Now the threshold $I$ represents
the peak intensity associated with the flare.  The rate of events with
threshold $I$, $R(I)$, is the number of events in the catalog with
peak intensity greater than $I$ divided by the total duration of the
catalog. Using the same thresholds as before for the whole catalog,
and rescaling the distributions using Eq.~(\ref{eq:2}) we obtain the
results shown in Fig.~\ref{fig:Pw_resc} (upper group of data).  In
this case power law behavior with an exponent $\gamma=2.19(5)$
($x^*\approx 0.28$) is observed at late times, although at short times
there are deviations from data collapse at lower thresholds.  The
deviations may be due to undercounting short waiting times following a
large flare, an obscuration effect previously pointed out by Wheatland
\cite{wheatland_2001}.  All of the data sets can also be fitted with
L\'{e}vy functions, which turn out to give comparable results except for
the waiting times at solar maximum, where the fit with L\'{e}vy
distributions is inferior~\cite{hans}. We choose here to focus on the fit
using a single function (Eq.~1) since it appears to describe all the
data sets equally well.

 Wheatland~\cite{wheatland_2001,Wheatland:Waiting,
wheatland02:_minmax} has modeled the solar flare waiting time
distribution in terms of a time-dependent Poisson process with a
flaring rate $\lambda(t)$.  When the flaring
rate varies slowly over a waiting time, $t_w$, the distribution of
waiting times can be written as  
\beq P(t_w,I)=
\frac{1}{\bar \lambda_I}\int_0^{\infty}F_I(\lambda) \exp(-\lambda t_w)
\lambda^2 d\lambda \quad ,
\label{eq:wheatland}
\eeq
where the average flaring rate ${\bar \lambda_I}=
 \int_0^{\infty}F_I(\lambda) \lambda d\lambda$ and
 $F_I(\lambda)\Delta\lambda$ is the fraction of the time the rate to produce
flares exceeding intensity $I$ is
 within $\Delta\lambda$ of $\lambda$.   
The function we find to fit the data has
 $f_I(\lambda) =\lambda^2 F_I(\lambda) / \bar \lambda$ 
corresponding to the Gamma distribution: 
\beq 
f_I(\lambda) \sim 
 \Big({\frac {\lambda x^*}{R(I)}}\Big)^{\gamma -1}
 \exp\left(-\frac {\lambda x^*}{R(I)}\right) \label{eq:gamma} \quad . 
\eeq 

 A mathematical equivalence with
 superstatistics formulas of Beck and Cohen~\cite{beck_cohen} can be
 made by mapping  $\lambda \rightarrow \beta$, $t_w \rightarrow E$ and
 $f(\lambda) \rightarrow f(\beta)$.
No necessary physical connection is implied  in this
equivalence, since one could just as well think of a subordination
mechanism~\cite{feller_book} as being at the basis 
of Eq.~(\ref{eq:wheatland}). 
It is also worth remarking that several turbulent systems have been
analyzed recently using the superstatistics framework.
These include velocity differences in Taylor-Couette
flow~\cite{Swinney}, 
intermittency of the wind~\cite{rapisarda} or solar 
wind~\cite{voros}.

At solar maximum the critical exponent $\gamma$ for the waiting time
distribution is close to $\gamma=3$.  This
implies that distribution of flaring rates $F_I(\lambda)$ is
close to exponential for a range of  $I$. Since high intensity flares predominately arise from
active regions during solar maximum, the origin of this distribution could be investigated
by tracking the flaring rates of individual active regions.

 At solar minimum, the critical exponent
$\gamma\approx 3/2$ for a range of intensities $I$, implying that the
distribution of flare rates 
\beq 
F_I^{\min}(\lambda) \sim
\lambda^{-3/2}\exp\left(-\frac {\lambda x^*}{R(I)}\right)
\label{eq:sub_critical} \quad . 
\eeq
 This formula describes the probability distribution for the number of
offspring in a subcritical branching process~\cite{harris}. 

Alternatively, the marginal behavior in on-off intermittency also
gives a distribution of laminar times with an exponent
$\gamma=3/2$~\cite{on-off}, in very good agreement with the quiet time
distribution at solar minimum. In fact an intermittent on-off
dynamo~\cite{dynamo} has been used to describe the solar cycle and
long term records of solar activity such as grand minima.  Our results
lead us to speculate that such a dynamo operating in a
marginal state may also be able to capture the quiet times of flares
- excluding active regions. Self-organized criticality may provide a
mechanism for this dynamo to sustain itself in a marginal state. Active
regions, superimposed on this fluctuating state, may represent
plasma instabilities in the on-off dynamo, with their own emergent behavior.


\begin{thebibliography}{33}
\bibitem{note1} Our $\gamma_{\rm dur}$ corresponds 
to $1/(q-1)$ of Ref.~\cite{beck_cohen}.
\end{thebibliography}
\end{document}